\documentclass[aps, prx, superscriptaddress, reprint, showpacs]{revtex4-1}

\usepackage{amsfonts}
\usepackage{amsmath}
\usepackage{amssymb, bm, mathrsfs}
\usepackage{graphicx, epstopdf, color}
\usepackage{soul}
\usepackage{appendix}
\usepackage[colorlinks=true,citecolor=blue]{hyperref}

\begin{document}

\title{Universal Model for Electron Thermal-Field Emission from \\ Two-Dimensional Semimetals}

\author{L. K. Ang}
\email{Invited Speaker}
\email{ricky\_ang@sutd.edu.sg}
\affiliation{Science, Mathematics and Technology, Singapore University of Technology and Design, Singapore 487372}

\author{Yee Sin Ang}
\email{yeesin\_ang@sutd.edu.sg}
\affiliation{Science, Mathematics and Technology, Singapore University of Technology and Design, Singapore 487372}

\author{Ching Hua Lee}
\affiliation{Department of Physics, National University of Singapore, Singapore 117542}

\begin{abstract}
We present the theory of out-of-plane (or vertical) electron thermal-field emission from 2D semimetals. 
We show that the current-voltage-temperature characteristic is well-captured by a \emph{universal} scaling relation applicable for broad classes of 2D semimetals, including graphene and its few-layer, nodal point semimetal, Dirac semimetal at the verge of topological phase transition and nodal line semimetal.
Here an important consequence of the universal emission behavior is revealed: in contrast to the common expectation that band topology shall manifest differently in the physical observables, band topologies in two spatial dimension are \emph{indistinguishable} from each others and bear no special signature in the electron emission characteristics. 
Our findings represent the quantum extension of the universal semiclassical thermionic emission scaling law in 2D materials, and provide the theoretical foundations for the understanding of electron emission from cathode and charge interface transport for the design of 2D-material-based vacuum nanoelectronics. 

\end{abstract}

\maketitle

\section{Introduction}
Since the first field emission model or the well known Fowler-Nordheim (FN) law was formulated about a century ago \cite{FN}, it remains an active topic  to discover various materials as emitters, geometrical configuration, and different operating conditions for diode physics and high power microwaves \cite{PHP2008,p_zhang,JAP2021}.
It is expected that the traditional emission models are no longer valid for for quantum materials such as two-dimensional (2D) materials like graphene, transition-metal dichalcogenide (TMD) materials, and topological materials (like topological
insulators and semimetals) \cite{IEEE2022,InfoMat2021}.
In two-dimensional (2D) materials, the extreme confinement of electrons within atomic-scale thickness and the existence of unusual emergent fermions with exotic energy dispersions lead to the occurrence of unusual transport properties that are radically different from that of the three-dimensional bulk materials \cite{neto, sangwan}.
Klein tunneling \cite{katsnelson}, unconventional quantum Hall effect \cite{novoselov}, quantum spin Hall insulating phase \cite{kane}, valley contrasting transport \cite{d_xiao} and spontaneous berryogenesis \cite{j_song} are some of the most extraordinary transport phenomena discovered in 2D materials.

Compared to field emission, thermionic emission due to 2D materials have been studied more intensively \cite{ang3,ang1,PRAp2016,PRAp2015,trushin1,trushin2}
In the case of thermionic emission where electrons are emitted from a solid surface via semiclassical thermal excitation pathway, the existence of a universal current-temperature scaling law represents another fascinating transport manifestations of the reduced dimensionality of 2D materials, where 
the universal temperature scaling laws were obtained \cite{ang1}, signifying the breakdown of the century-old Richardson-Dushman thermionic emission theory \cite{RD}.

For field emission \cite{jensen_text}, electron can undergo quantum mechanical tunneling due the narrowing of the interface potential barrier caused by an external applied electric field. 
For classical materials, the Fowler-Nordheim (FN) theory \cite{FN, schottky} provides the key theoretical foundation, and has formed one of the central pillars of vacuum electronics at the dawn of twentieth century. Remarkably, field emission physics remains highly important to solid-state devices in the post vacuum tube era because of its critical role in the interfacial charge injection process across the metal/insulator and metal/semiconductor interfaces \cite{sze} which are omnipresent in modern electronic and optoelectronic devices.
Due to its technological importance in both vacuum \cite{han} and solid-state device technology, FN theory has been continually refined over the past decades \cite{MG, burgess, stratton, forbes2, jensen, kyritsakis, sd_liang, wei,Steve,Sim,Sim2}.

\begin{figure}[t]
	\includegraphics[scale=0.485]{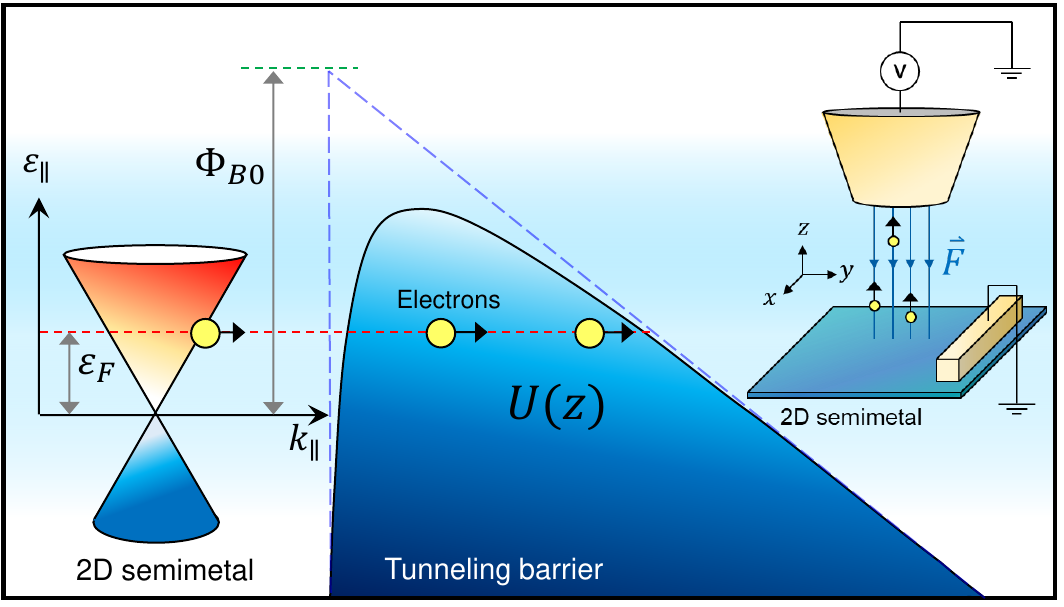}
	\caption{Band diagram of electron field emission from 2D semimetals. $U(z)$ is the interface tunneling potential barrier inclusive of the image charge. Inset: Schematic drawing of electron field emission setup with an applied electric field $F$. }
\end{figure}
 
For quantum 2D materials, the validity of FN based models need to be scrutinized since the foundational bases of such theories, particularly the assumption of 3D quasi-free electron gas with parabolic dispersion, are fundamentally incompatible with the physical properties of 2D material \cite{ang2}. 
Despite ongoing experimental \cite{eda, qian, wu, xiao, palnitka, yamaguchi, huang, moussa, wu_GVT, wu_SFE, xu, bartolomeo, rezaeifar, q_ma} and theoretical \cite{wei3, song, wang, d_jena} efforts in unearthing the physics of electron emission from graphene, a comprehensive understanding of electron out-of-plane (or vertical) field emission from the broader family of 2D semimetals remains largely incomplete thus far. 
The following questions remain open: 
What is the radical difference(s) between the electron field emission in 2D and 3D materials? Is there a simple unifying scaling relation, counterpart to the universal current-temperature scaling law of semiclassical thermionic emission \cite{ang1}, that encompasses the field emission characteristics for broad classes of 2D semimetals? 
Recently, we have reported that the FN law is no longer valid for 3D Dirac and Weyl semi-metals \cite{PRB2021}.
These non-FN field emission models will not converge to space charge limited current condition \cite{PRAp2021}.

In this work, we will address the above questions by developing the theory of out-of-plane electron thermal-field emission from 2D semimetals. We present the full numerical and analytical approximate expressions of the thermal-field emission current \cite{MG} for various 2D semimetals, including graphene and its few-layer \cite{neto}, nodal point semimetals \cite{burkov} of generalized pseudospin vorticity \cite{moore}, Dirac semimetals at the verge of topological quantum phase transition \cite{roy} and nodal line semimetals \cite{fang}. 
Remarkably, by employing a generalized model of 2D semimetals \cite{ang1}, we show that the vertical emission current density ($\mathcal{J}_{\text{2D}}$) for broad classes of 2D semimetals can be accurately captured by a \emph{universal} current-field-temperature scaling law,
\begin{equation}\label{uni0}
\mathcal{J}_{\text{2D}} (F, T) \approx \mathcal{A} \frac{\pi d_F}{c \sin\left( \pi /c\right)}\exp\left( -\frac{ \mathcal{B} }{ F } \right), 
\end{equation}
where $\mathcal{A}$ and $\mathcal{B}$ are material-dependent parameters, $F$ is the electric field strength, $T$ is the emitter temperature, $d_F \propto F$ is a field-dependent parameter (defined below), and $c \equiv d_F/k_BT$.
A particularly intriguing feature of Eq. (\ref{uni0}) is its departure from the bulk material thermal-field emission scaling relation of $\mathcal{J}_{\text{3D}} = \mathcal{A} \pi d_F^2 / c\sin\left(\pi/c\right) \exp\left( -\mathcal{B} / F \right)$ \cite{MG}, thus revealing the breakdown of the classic FN-type electron emission model.
Note this linear dependence of the field in the prefactor (for field emission) is similar to the pure thermionic emission reported \cite{ang1}.

The universal field emission scaling law identified here reveals an important fundamental feature of band topology in two spatial dimensions: in contrast to the common expectation that different band topologies shall manifest differently in the electronic, optical, electrical, mechanical and thermodynamical properties, as already demonstrated in the density of states \cite{j_wang}, transport \cite{j_hu, hp_sun}, optical response \cite{tabert,carbotte}, quantum oscillations \cite{c_wang, c_li}, many-body physics \cite{roy2, sur}, and shear viscosity \cite{moore}, band topologies in two spatial dimensions are \emph{indistinguishable} from each others and bears no special signatures in the electron emission current-field scaling characteristics. 
Nonetheless, the \emph{Fermi level dependence} of the emission current does reflects the differences in the energy dispersion of various 2D semimetals.
As electron field emission is a crucially important charge transport mechanism in both solid/vacuum \cite{p_zhang} and solid/solid interfaces \cite{allain, xu_rev}, the model developed here shall offer a pivotal theoretical foundation for both the fundamental understanding of surface and nanoscale interface physics, and the practical design of high-performance vacuum \cite{zhou} and solid-state devices based on 2D materials and their van der Waals heterostructures \cite{am_liang} and semi-metal based electrical contact \cite{Li2021,PRAp2020,PRAp2020b}

\section{Field emission model for 2D materials}

In a 2D system, the confinement of electrons within the 2D plane lead to the formation of discrete subbands. 
The energy dispersion and the wavevector of the $i$th-subband are, $\varepsilon_{\mathbf{k}_i} = \varepsilon_{\parallel, i} (\mathbf{k}_{\parallel, i}) + \varepsilon_{\perp, i} (\mathbf{k}_{\perp, i})$ and $\mathbf{k}_i = \mathbf{k}_{\parallel, i} + \mathbf{k}_{\perp, i}$, respectively, where $\varepsilon_{\perp, i}$ and $\mathbf{k}_{\perp, i}$ denotes the discrete subband energy and the quantized wavevector along the out-of-plane direction, respectively; $\varepsilon_{\parallel,i}$ and $\mathbf{k}_{\parallel,i}$ denotes the continuously-dispersing \emph{in-plane} energy dispersion and wavevector, respectively.
The out-of-plane electron emission current density can be generally expressed as \cite{moore}:
\begin{equation}\label{J_2D}
\mathcal{J}_{\text{2D}}(F,T) = \frac{ge}{\left( 2\pi \right)^2 } \sum_i \tau_{\perp, i}^{-1} \int_{\text{B.Z.}} d^2 \mathbf{k}_{\parallel, i} D(\mathbf{k}_i, F) f(\mathbf{k}_i, \mathbf{k}_F) ,
\end{equation}
where $g$ is the degeneracy factor, $\mathbf{k}_F$ is the Fermi wavevector, $D(\mathbf{k})$ is the \emph{out-of-plane} tunneling probability, $f(\mathbf{k}_i)$ is the Fermi-Dirac distribution function, the $\mathbf{k}_\parallel$-integral spans the whole Brillouin zone, $\sum_i$ runs over all subbands, and $\tau_\perp^{-1}$ is the vertical electron injection rate, which is affected by the intrinsic material properties and the device configuration, and can be experimentally extracted from the transport measurements \cite{xia, sinha, massicotte}. 
The transmission probability, $D(\mathbf{k}_i, F)$, is a function of the \emph{total} wavevector, $\mathbf{k}_i$, instead of only the out-of-plane component $\mathbf{k}_{\perp,i}$, due to the $\mathbf{k}_{\parallel, i}$-nonconserving scatterings \cite{meshkov, russell, zhu, britnell_NC, britnell_SC, t_roy, ang3}, which can arise extrinsically from electron-electron \cite{meshkov}, phonon \cite{vdovin}, magnon \cite{ghazaryan}, and defects \cite{russell, chandni, trushin1} scatterings, and intrinsically from the uncertainty principle as a result of confining electrons within the 2D plane \cite{trushin2}. 
Such momentum nonconservation leads to the coupling of the $\mathbf{k}_{\perp, i}$ and $\mathbf{k}_{\parallel, i}$ during the out-of-plane tunneling process. 

%
%
 
For field-induced electron emission, the tunneling potential barrier is modeled, with inclusion of the image charge effect across a dielectric interface, as $U(z) = \Phi_{B0} - eFz - e^2\nu/16\pi\epsilon_0 z$, where $\Phi_{B0}$ is the \emph{intrinsic} interface potential barrier height measured from zero-energy, $\nu \equiv \left(\epsilon_s - 1 \right) / (\epsilon_s + 1)$, and $\epsilon_s$ is the dielectric constant of the substrate material (see Fig. 1 for the band diagram). 
The corresponding transmission probability can be accurately approximated as \cite{MG}, 
%
%
$D(\mathbf{k}_i, F) \approx D_F \exp \left\{ \left[\varepsilon_\parallel(\mathbf{k}_{\parallel,i}) + \varepsilon_{\perp,i}(\mathbf{k}_{\perp, i}) - \varepsilon_F \right] / d_F \right\}$, where $d_F \equiv \hbar e F / [(8m)^{1/2} \left(\Phi_{B}-\varepsilon_F\right)^{1/2} t_0]$, $D_F \equiv \exp\left[-[4 (2m)^{1/2} \left(\Phi_{B0} - \varepsilon_F \right)^{3/2} v_0] /3\hbar e F\right]$, $v_0$ and $t_0$ are correction factors for the image charge effect, and $m$ is the free electron mass (see \cite{SM} for detailed derivations). 
%
%
%
Consider the typical case where there is only one subband around the Fermi level $\varepsilon_F$, Eq. (\ref{J_2D}) becomes, 
\begin{equation}\label{J_2D_3}
\mathcal{J}_{\text{2D}} (F, T) = \frac{ ge D_F }{ 4\pi^2\tau_\perp } \int_{\text{B.Z.}} \text{d}^2\mathbf{k}_\parallel \frac{ \exp\left( \frac{\varepsilon_\parallel - \varepsilon_F}{d_F} \right)  }{ \exp\left( \frac{\varepsilon_\parallel - \varepsilon_F}{k_BT} \right) + 1  },
\end{equation}
where $k_\parallel$ and $\varepsilon_\parallel$ denote the in-plane wavevector and energy components of the subband undergoing emission. 

We now employ a generalized 2D anisotropic density of states (DOS) to solve Eq. (\ref{J_2D_3}). 
The DOS is defined as, $\rho(\varepsilon_\parallel) \text{d} \varepsilon_\parallel = \left[ g/(2\pi)^2 \right] \int_0^{2\pi} \text{d}\phi k_\parallel \text{d}  k_\parallel $ where $k_\parallel \equiv \left| \mathbf{k}_\parallel \right|$. 
Very generally, we can express the $k_\parallel$-differential as a two-variable polynomials, i.e. $k_\parallel \text{d}  k_\parallel = \sum_{m,1 = 0}^{\infty} \beta_{ml}  \left| \varepsilon_\parallel \right|^m \cos^l\phi \text{d}\varepsilon_\parallel$ where $\beta_{ml}$ is the expansion coefficient of the $m$-th order in $\varepsilon_\parallel$ and $l$-th order in $\cos\phi$, which yields the \emph{generalized 2D anisotropic DOS}, 
\begin{equation} \label{uni_DOS}
\rho(\varepsilon_\parallel) = g\sum_{m, l =0}^{\infty} \bar{\beta}_{ml} \left|\varepsilon_\parallel \right|^m, 
\end{equation}
where $m \in \mathbb{Z}^{\geq0}$ and $\bar{\beta}_{ml}$ is an expansion coefficient that compactly contains the $\phi$-integral \cite{SM}. Combining Eqs. (\ref{J_2D_3}) and (\ref{uni_DOS}), the \emph{generalized} thermal-field emission electrical current density from a 2D semimetal becomes $\mathcal{J}_{\text{uni}} = \sum_{m,l}^\infty \mathcal{J}_{\text{uni}}^{(m,l)}$ where the $(m,l)$-component is
\begin{equation}\label{J_uni}
\mathcal{J}_{\text{uni}}^{(m,l)} = \frac{ge D_F}{\tau_\perp}  \bar{\beta}_{ml} \int_{-\infty}^{\Phi_{B0}} d\varepsilon_\parallel  \left|\varepsilon_\parallel\right|^m \mathcal{G}_{\text{TF}},
\end{equation}
which cannot be analytically solved except for the case of $m=0$. 
An approximate \emph{analytical} expression can be obtained by making the following \emph{ansatz}. 
The term $\left|\varepsilon_\parallel\right|^m$ is replaced by $\xi_m \left|\varepsilon_F\right|^m $ where $\xi_m$ is a the correction factor for the $m$-th order term since field emission is dominantly contributed by electrons residing around $\varepsilon_F$. 
By making the substitution $u = \exp\left[\left(\varepsilon_\parallel - \varepsilon_F\right)/ d_F\right]$, Eq. (\ref{J_uni}) becomes \cite{SM}
\begin{equation}\label{uni1}
\mathcal{J}_{\text{uni}}^{(m,l)} \approx \xi_m \frac{eD_F}{\tau_\perp}  \bar{\beta}_{ml} \left|\varepsilon_F\right|^m d_F \int_0^{\infty} \frac{d \text{d} u}{ u^c + 1 },
\end{equation}
where $c \equiv d_F / k_BT$. The above integral can be analytically solved \cite{MG} for $c>1$ t,
\begin{equation}\label{uni2}
\mathcal{J}_{\text{uni}}^{(m,l)} \approx \xi_m \frac{e}{\tau_\perp} \sum_{m,l = 0}^\infty \bar{\beta}_{ml} \left|\varepsilon_F\right|^m \frac{\pi d_F}{c\sin\left( \pi/c \right)} D_F.
\end{equation}
We evaluate the full numerical solution in Eq. (\ref{J_uni}) and the analytical approximation in Eq. (\ref{uni2}) for $m=0$ to $m=4$ (Fig. 2). The approximate solution is in agreement with the full numerical results over the range of $F = 1$ V/nm to $F = 5$ V/nm with $\xi_m = (1, 1.22, 4, 24, 220)$ for $m = 0$ to $m=4$. 
From Eq. (\ref{uni2}), we obtain the following scaling law, 
\begin{equation} \label{uni}
\mathcal{J} \propto \frac{\pi d_F}{c\sin\left( \pi/c \right)} D_F.
\end{equation}
Equation (\ref{uni}) represents a \emph{universal} current-field-temperature scaling law for 2D semimetals as long as the DOS near $\varepsilon_F$ can be captured by the generalized DOS model in Eq. (\ref{uni_DOS}). 
Equation (\ref{uni}) thus offer a simple unifying scaling law description of the field emission characteristics in two spatial dimensions.
We derive the field emission characteristics for several representative 2D semimetals, including: (i) nodal point semimetal; (ii) graphene and its few-layer; (iii) Dirac semimetal near topological quantum phase transition; and (iv) nodal line semimetal, and show that their current-field-temperature scaling relation universally converges to Eq. (\ref{uni}) (summarized in Fig. 2).
\begin{figure*}
	\includegraphics[scale=0.76858]{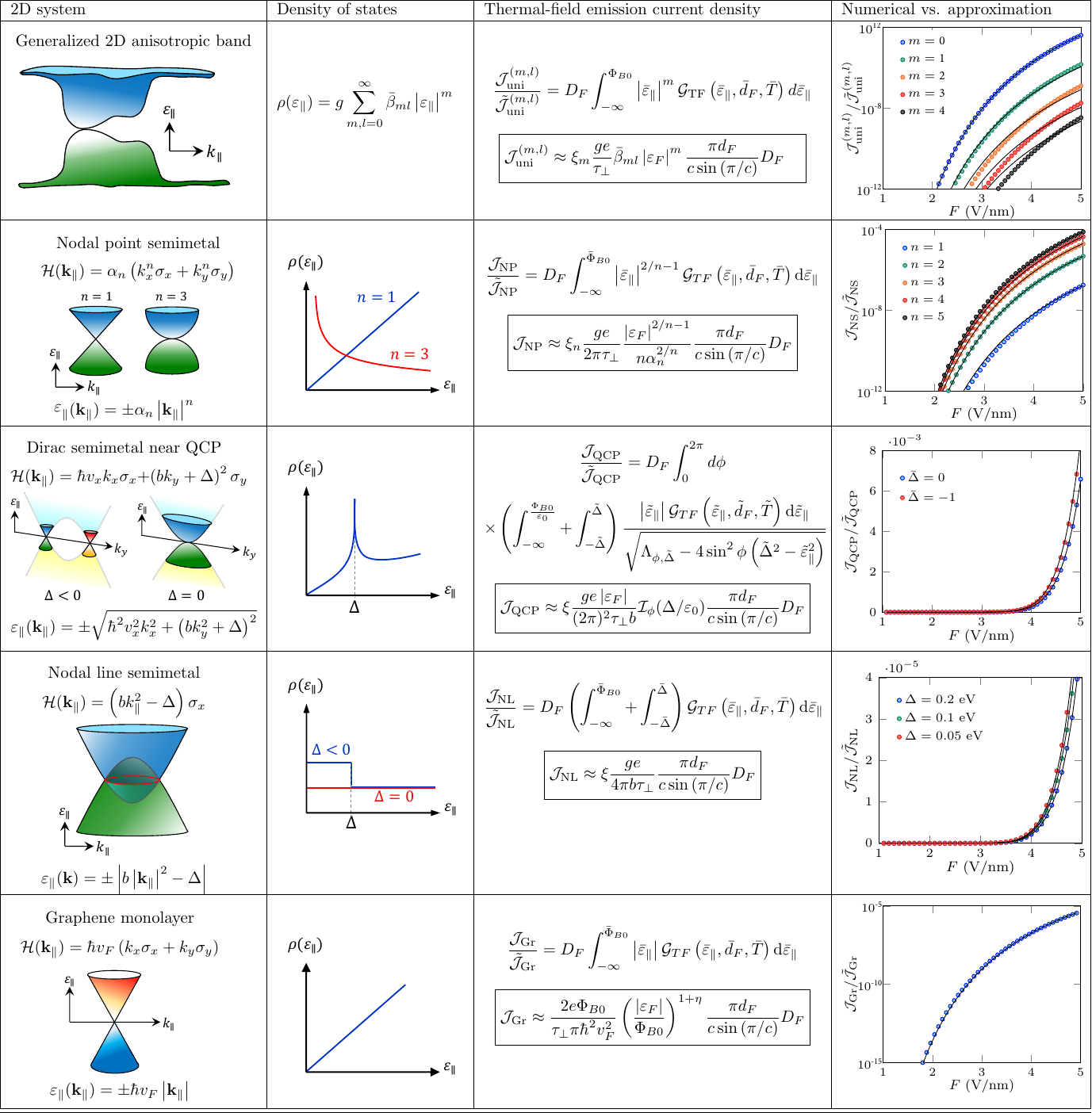}
	\caption{Thermal-field electron emission from 2D semimetals. Energy dispersion(column 1), density of states (column2), full numerical and the approximate expressions of the thermal-field emission current density (column 3), and the corresponding dimensionless forms (column 4) evaluated with $T = 300$ K, $\epsilon_s = 3.5$. $\varepsilon_F = 0.1$ eV, and $\Phi_{B0}=4.5$ eV. The full numerical and the analytical approximate solution are denoted by circle and solid curves, respectively. \textcolor{black}{The normalization factors of the current densities (i.e. $\tilde{\mathcal{J}}_{\text{uni}}^{(m,l)}$, $\tilde{\mathcal{J}}_{\text{NP}}$, $\tilde{\mathcal{J}}_{\text{QCP}}$, and $\tilde{\mathcal{J}}_{\text{NL}}$, $\tilde{\mathcal{J}}_{\text{Gr}}$) are listed in Table I.} }
	\end{figure*}

\section{Case Studies of different 2D semimetals}

\textcolor{blue}{\textbf{\emph{Nodal point semimetal.}}}-- We first consider a generalized model of topological 2D nodal point semimetal with the effective Hamiltonian, 
\begin{equation}\label{H_NP}
\hat{\mathcal{H}}_{\text{NP}}(\mathbf{k}_\parallel) = \alpha_n \left(k_x^n \sigma_x + k_y^n \sigma_y\right),
\end{equation}
where $\mathbf{k}_\parallel = (k_x, k_y)$ is the in-plane wavevector, $\sigma_{x,y}$ is the Pauli matrices, $\alpha_n$ is a material-dependent parameter, and $n \in \mathbb{Z}^+$ denotes the pseudospin vorticity.
Equation (\ref{H_NP}) also describes the low-energy quasiparticles in $n$-layer graphene of Bernal stacking configuration with $\alpha_n = (\hbar v_F)^n/t_\perp^{n-1}$, $v_F = 10^6$ m/s and $t_\perp = 0.39$ eV \cite{neto}.
Using the same procedure described above, Eq. (\ref{J_2D_3}) can be approximately solved as
\begin{equation}\label{J_NP}
\mathcal{J}_{\text{NP}} \approx \xi_n \frac{ge}{ 2\pi \tau_{\perp} } \frac{\left|\varepsilon_F\right|^{2/n - 1} }{n \alpha_n^{2/n}} \frac{\pi d_F}{c\sin\left(\pi/c\right)} D_F
\end{equation}
The analytical approximate solution in Eq. (\ref{J_NP}) with the correction factor, $\xi_n = (1.58, 1, 1.2, 1.43, 1.65)$ for $n=1$ to $n=5$ based on the few-layer graphene band parameters exhibits good agreement with the full numerical results (Fig. 2). 

\textcolor{blue}{\textbf{\emph{Dirac semimetal near topological QCP.}}}-- We next consider a 2D Dirac semimetal near the topological quantum critical point (QCP) which can be captured by the effective Hamiltonian,
\begin{equation}\label{H_QCP}
\hat{\mathcal{H}}_{\text{QCP}}(\mathbf{k}_\parallel)  = \hbar v_x k_x \sigma_x + (bk_y^2 + \Delta) \sigma_y ,
\end{equation}
where $v_x$ and $b$ are band structure parameters. The system undergoes a topological quantum phase transition from a Dirac semimetal into a band insulator when $\Delta$ is switched from $\Delta \leq 0 $ to $\Delta \geq 0$. In the semimetallic phase ($\Delta \leq 0$), the thermal-field emission current density ($c>1$) is
\begin{equation}\label{J_QCP}
\mathcal{J}_{\text{QCP}} \approx \xi\frac{ge\left|\varepsilon_F\right| }{(2\pi)^2 \tau_\perp b} \mathcal{I}_{\phi,\varepsilon_F}(\Delta/\varepsilon_0) \frac{ \pi d_F }{ c \sin\left(\pi/c\right) } D_F
\end{equation}
where $\varepsilon_0 \equiv \hbar^2 v_F^2 / b$ is a characteristic energy, and $\phi = \tan^{-1} \left( k_y/k_x \right)$. The $\phi$-integral,  $\mathcal{I}_{\phi,\varepsilon_F} (\mu) \equiv \int_0^{2\pi} \text{d}\phi \left[ \Lambda_{\phi, \mu} - 4\sin^2\phi\left( \mu^2 - \varepsilon_F^2/\varepsilon_0^2 \right) \right]^{1/2} $ where $\Lambda_{\phi, \mu} \equiv (\cos^2\phi + 2\mu\sin^2\phi)^2$, arises from the broken rotational symmetry of Eq. (\ref{H_QCP}). The approximate solution Eq. (\ref{J_QCP}) again exhibits good agreement with the full numerical results (Fig. 2), with $\xi = 1.35$ at the QCP (i.e. $\Delta = 0$) and $\xi = 2.46$ at the semimetallic phase (i.e. $\Delta = 0.1$ eV). 

\textcolor{blue}{\textbf{\emph{Nodal line semimetal.}}}-- In a 2D nodal line semimetal, the band touching of two bands extends from discrete nodal point into a continuous one-dimensional opened nodal line or closed nodal ring in phase space. We consider a representative 2D semimetal that hosts an isotropic equienergy nodal loop as described by the effective Hamiltonian,
\begin{equation}\label{H_NL}
\mathcal{H}(\mathbf{k}_\parallel) = \left(b k_\parallel^2 - \Delta \right) \sigma_x 
\end{equation}
where $\Delta$ is a band inversion parameter. Solving Eq. (\ref{J_2D_3}) yields the approximate solution,
\begin{equation}\label{J_NL}
\mathcal{J}_{\text{NL}} \approx \xi\frac{g e }{4\pi b \tau_\perp} \frac{\pi d_F}{c\sin\left( \pi/c \right)} D_F.
\end{equation}
The above approximated form produces good agreement with the full numerical evaluation of Eq. (\ref{J_2D_3}) (Fig. 2) with the correction factor, $\xi = (1.27, 1.53, 1.72)$ for $\Delta = (0.05, 0.1, 0.2)$ eV.

\textcolor{blue}{\textbf{\emph{Graphene.}}}-- The thermal-field emission from graphene is described by the Dirac Hamiltonian, $\mathcal{H}(\mathbf{k}_\parallel) = \hbar v_F\left( k_x \sigma_x + k_y \sigma_y \right)$. The field emission current density is obtained as,
\begin{equation}\label{J_Gr}
\mathcal{J}_{\text{Gr}} \approx \frac{2e \Phi_{B0} }{ \tau_\perp \pi \hbar^2 v_F^2}  \left( \frac{ \left|\varepsilon_F\right| }{ \Phi_{B0}} \right)^{1+\eta} \frac{ \pi d_F }{ c\sin\left(\pi/c\right)} D_F.
\end{equation}
Here, the $\varepsilon_F$ (also implicit in $d_F$, $c$ and $D_F$) is a function of the applied electric field strength, i.e. $\varepsilon_F = \varepsilon_F(F)$ \cite{yu_wf}, due to the field-effect tunable Fermi level in graphene. 
Assuming a planar geometry and an initially undoped graphene with $\varepsilon_F \approx 0$, we obtain $\varepsilon_F(F) = \sqrt{\pi\hbar^2v_F^2\varepsilon_0F/e}$. By evaluating Eq. (\ref{J_Gr}) with $\varepsilon_F(F)$, the full numerical solution is accurately reproduced with $\eta = 3/17$.

\begin{table}[t]
	\centering
	\caption{\textcolor{black}{Definition of the current density normalization factors ($\bar{\mathcal{J}}$). Here, $\bar{\varepsilon}_\parallel \equiv \varepsilon_\parallel / \Phi_{B0}$, $\bar{d}_F \equiv d_F / \Phi_{B0}$, $\bar{T} \equiv k_BT / \Phi_{B0}$, $\tilde{\varepsilon}_\parallel \equiv \varepsilon_\parallel / \Phi_{B0}$, $\tilde{d}_F \equiv d_F / \Phi_{B0}$, $\tilde{T} \equiv k_BT / \Phi_{B0}$, and $\tilde{\Delta} \equiv \Delta/\varepsilon_0$.}  }
	\begin{tabular}{ l l l } 
		\hline \hline 
		2D System && $\bar{\mathcal{J}}$ \\
		\hline 
        Generalized 2D band && $\tilde{\mathcal{J}}_{\text{uni}}^{(m,l)} = ge\tau^{-1}_{\perp} \sum_{m,l =0}^{\infty} \bar{\beta}_{ml }$ \\
		Nodal point semimetal && $\tilde{\mathcal{J}}_{\text{NP}} = \frac{ge \Phi_{B0}^{2/n} }{ 2\pi n\tau_{\perp} \alpha_n^{2/n} }$ \\ 
		Dirac semimetal near QCP && $\tilde{\mathcal{J}}_{\text{QCP}} = \frac{\xi ge \varepsilon_F \Phi_{B0} \mathcal{I}_{\phi,\varepsilon_F}(\Delta/\varepsilon_0)}{4\pi^2\tau_\perp b} $ \\
		Nodal line semimetal && $\tilde{\mathcal{J}}_{\text{NL}} = \frac{ge \Phi_{B0}}{4\pi \tau_\perp b}$ \\
		Graphene && $\tilde{\mathcal{J}}_{\text{Gr}} = \frac{2e \Phi_{B0}^2}{ \pi \tau_\perp \hbar^2 v_F^2 } $ \\
		\hline\hline
	\end{tabular}
\end{table}

\begin{table}[t]
	\centering
	\caption{ The current-Fermi-level scaling relation for various 2D semimetals. The normalization factor is defined as $\psi \equiv \pi d_FD_F/c \sin\left( \pi/c \right)$. }
	\begin{tabular}{ l l c } 
		\hline \hline 
		2D System && Fermi level dependence of $\mathcal{J} / \psi $ \\
		\hline 
		Nodal point semimetal && $ \left| \varepsilon_F \right|^{2/n - 1} $ \\ 
		Dirac semimetal near QCP && $ \left| \varepsilon_F \right| \mathcal{I}_{\phi,\varepsilon_F}(\Delta/\varepsilon_0) $ \\
		Nodal line semimetal && $ \text{constant in }\varepsilon_F$ \\
		Graphene && $ \left| \varepsilon_F \right|^{1+\eta}$ \\
		\hline\hline
	\end{tabular}
\end{table}

\section{Discussions and Conclusions}

Before conclusions, we make \textcolor{black}{six} remarks. Firstly, in the `cold' field emission regime of $T \to 0$, the scaling law in Eq. (\ref{uni}) becomes $\mathcal{J}_{\text{2D}}(F, T \to 0) \propto d_F D_F$, which is in stark contrast to the FN-type cold field emission from bulk materials, $\mathcal{J}_{\text{3D}}(F, T \to 0) \propto d_F^2 D_F$. 
This difference directly suggests the breakdown of the FN model in 2D semimetals. 

Secondly, the universal scaling law breaks down in the case of a \emph{non-dispersing} 2D flat bands in 2D systems such as the Kagome \cite{mielke}, Lieb \cite{lieb}, $\alpha$-$T^3$ \cite{raoux} and Archimedian \cite{lima} lattices.
For a flat band situated at $\varepsilon_\parallel = \varepsilon_{\text{FB}}$, the thermal-field emission current densities is
\begin{equation}
\mathcal{J}_{\text{FB}}\propto D_F \frac{\exp\left( \frac{\varepsilon_{\text{FB}} - \varepsilon_F}{d_F} \right)}{ 1 + \exp\left(\frac{\varepsilon_{\text{FB}} - \varepsilon_F }{k_BT}  \right) },
\end{equation}
which clearly deviates from Eq. (\ref{uni}). 
Such deviation also explains the reduced accuracy of Eq. (\ref{uni2}) in approximating $\mathcal{J}_{\text{uni}}^{(m,l)}$ using Eq. (\ref{J_uni}) at large $m$ since each $\left|\varepsilon\right|^m$ term in Eq. (\ref{uni_DOS}) corresponds to $\varepsilon_\parallel\propto k_\parallel^{2/(m+1)}$ and thus a larger $m$ corresponds to a `flatter' dispersion.

Thirdly, we remark that although the $\mathcal{J}$-$F$ scaling does not offer distinctive signature of band topology, the $\mathcal{J}$-$\varepsilon_F$ scaling does contain rich scaling signatures for different band topologies (see Table I), which may offer a tool to probe the band topology in 2D systems.

\textcolor{black}{
Fourthly, the injection time parameter ($\tau_\perp$) of a 2D material solid/solid interface ranges from 10 ps (low-quality contact) to 0.1 ps (high-quality contact) depending on the contact quality \cite{sinha}. For electron emission into vacuum, although no direct experimental measured values are available thus far, we expect $\tau_\perp$ to be larger than the 0.1 ps in 2D-material/vacuum interface since 2D material surface readily ‘couples’ to vacuum without additional interfacial defect generation. 
This aspect is unlike the case of solid/solid interfaces in which the fabrication process inevitably introduces interface defects that could reduce the charge injection efficiency. Importantly, our theoretical model shall provide a route for the experimentalist to additionally extract the $\tau_\perp$ parameter from the experimental current measurement data. The $\tau_\perp$ parameter shall thus provide an additional term for characterizing the performance and quality of a 2D-material-based field emitter.}

\textcolor{black}{
Fifthly, we remark that the substrate can significantly affect the field emission characteristics of 2D field emitter, particularly the electronic properties of atomically-thin materials can be sensitively modified when they are in close proximity to an external materials \cite{vdw}. Here we briefly discuss how various substrate are expected to affect the electron emission of 2D field emitters. For suspended 2D field emitter, the solid/solid interface defects are expected to be minimal due to the absence of a supporting substrate. In this case, the shorter injection time parameter $\tau_\perp$ is expected to lead to an improved magnitude of the field emission current in such emitter. Conversely for 2D field emitter lying on a dielectric substrate, the presence of a solid/solid supporting surface could lead to larger amount of interface defects which could reduce the charge injection efficiency into vacuum. For 2D materials lying on a 3D metal substrate, the field emission current is expected to be dominated by the 3D bulk metal as the electronic density of states of 3D metals are typically much larger than that of metallic 2D materials. Here, instead of being an electron emission source, 2D materials serve as a surface coating layer that modulates the 3D bulk metal surface properties, which could significantly modify the field emission properties of the 3D bulk metal \cite{jensen_DFT}. We thus anticipate the substrate effects in 2D field emitter to form an important topic for future investigations.
}
%

\textcolor{black}{
Finally, we briefly comment on the edge emission from 2D materials, which is expected to differ significantly from the ‘face emission’ reported in this manuscript. The emission current density from edge is a 1D linear current density whose underlying transport physics and formalism is completely different from Eq. (2). Furthermore, the atomically sharp edge may generate strong field enhancement \cite{field} and image charge effect that is drastically different from the case of a planar surface \cite{image}. Importantly, the emitted electrons can also acquire a velocity component transversal to the edge emission direction. Such transversal electron emission can further complicate the formalism of edge emission. Microscopically, different edge termination configurations can affect the electronic properties of 2D materials \cite{edge}. The edge emission characteristics is thus expected to be sensitively influenced by the atomic profile of the 2D material edges. The theoretical modeling of edge emission from 2D materials are thus expected to be a more complex problem that involves a multiphysics approach of first-principle atomistic material simulations, electrostatic modeling, and charge injection theory.
}

In summary, we have developed the theory of out-of-plane thermal-field electron emission from 2D semimetals. 
We demonstrated the existence of a universal current-field scaling law broadly applicable for a large variety of 2D semimetals with different band topologies. 
As thermal-field emission represents one of the key charge transport process across solid/vacuum and solid/solid interfaces, the universal scaling law developed here shall provide a simple useful theoretical tool for the study and the design of vacuum electronics, nanoelectronics, optoelectronics and the emerging concepts of spintronic \cite{roche}, valleytronic \cite{ang4} and neuromorphic \cite{l_sun} devices based on 2D materials and their van der Waals heterostructures, and shall offer a theoretical basis for the understanding of complex electron emission phenomena such as ultrashoft-pulsed laser-induced internal photoemission \cite{heide} and photo-assisted hot carrier tunneling \cite{rezaeifar} in the 2D Flatland.

\section{acknowledgments}
This work is supported by A*STAR AME IRG (A2083c0057). 
Y.S.A is supported by the Singapore Ministry of Education Academic
Research Fund Tier 2 (MOE-T2EP50221-0019).

\end{document}